# Magneto-thermomechanically triggered active mechanical metamaterials – untethered, reversible, reprogrammable transformations with shape locking

Bihui Zou[1], Zihe Liang[1], Zhiming Cui[1], Kai Xiao[1], Shuang Shao[1], Jaehyung Ju[1]*

1. UM-SJTU Joint Institute, Shanghai Jiao Tong University, 800 Dongchuan Road, Shanghai 200240, China

*Corresponding author. Email: jaehyung.ju@sjtu.edu.cn

**Abstract**

Future active metamaterials for reconfigurable structural applications require fast, untethered, reversible, and reprogrammable (multimodal) transformability with shape locking. Herein, we aim to construct and demonstrate a magneto-thermomechanical tool that enables a single material system to transform with untethered, reversible, low-powered reprogrammable deformations and shape locking via the application of magneto-thermomechanically triggered prestress on a shape memory polymer and structural instability with asymmetric magnetic torque. We demonstrate the mutual assistance of two physics concepts — magnetic control combined with the thermomechanical behavior of shape memory polymers, without requiring new materials synthesis and high-power energy for reprogramming. Our approach can open a new path of active metamaterials, flexible yet stiff soft robots, and multimodal morphing structures, where we can design them in reversible and reprogrammable ways.

## Introduction

Future structural materials are evolving as robotic matter [1-5]; A material itself functions as a robotic machine that can have i) ultra-variable stiffness – both soft and stiff, ii) reversible transformation from one shape to others, and iii) reprogrammability. The multidisciplinary research community is actively securing this opportunity from rationally designed structures (metamaterials) with a combination of intelligent materials. For ultra-variable stiffness, a material must possess a dramatic stiffness transition of up to two orders of magnitude: low stiffness for large deployment and high stiffness for load carrying. The innovative transformation requires fast, untethered, and reversible deformations. The reprogrammability requires a multimodal transformation to change the anisotropy without replacing the original material.

Active metamaterials with magnetic control have recently shown great potential for reconfigurability with multimodal deformations achieved via microrotation of magnetic particles embedded in elastomers in a magnetic field [1,6-13]. This method has a unique advantage for future active metamaterials because of the fast and remote control of the shape transformation. However, an external magnetic field must be applied to retain the deformed shape for tunable structural applications because of the lack of shape-holding[14]; once the magnetic field is removed, the deformed structures recover their original shapes.

For shape locking, thermomechanical control of shape memory polymers (**SMPs**) has a significant advantage over other triggering methods; the molecules of SMPs freeze with the removal of external heat [15-18]. For example, a transformation of an amorphous SMP is locked after cooling below the glass transition temperature, $T_g$. Upon heating back above $T_g$, the locked shape is transformed back into the original stress-free shape. However, despite the shape-locking property, the transformations of SMPs are intrinsically irreversible [19,20]. Some groups have used liquid crystalline elastomers (LCEs)[2,21] to achieve reversible transformations. However, the complex material synthesis methods[22,23] and relatively low stiffness without shape locking[2] restrict their application for the robotic matter.

In parallel, multimodal deformation with reprogrammability is significant for future robotic matter because of its adaptable capability of a single material system for multi-inputs[6,24]. To date, two reprogramming methods have been dominant: structural design and materials synthesis. In the structural design, the asymmetric arrangement of self-contacted beams can produce a simple bending and folding mode deformation[6]. On the reprogrammability on the material level, de- and re-magnetization of low-Curie-temperature magnetic materials such as chromium dioxide ($CrO_2$) by laser heating is one method[11,25]. Heating a phase-change material above its melting temperature, $T_m$, to reorient encapsulated magnetic particles is another reprogramming method[26]. The assembly of magnetic soft-material modules by welding demonstrated a new reprogrammability[27]. However, these methods all require extra high-power thermal energy — laser and welding for reprogramming and a long reprogramming time [24].

Recently, a thermomechanical model allowed a single material system to transform with reversible and multimodal (reprogrammable) deformations[28]. With transformation aids for local prestress control and a temperature-dependent reverse stiffness effect, SMP structures could reversibly transform with shape locking in a multimodal manner. However, despite the breakthrough, the transformation aids need to assemble and disassemble with the SMP lattices for each reconfiguration cycle, which may not be considered a fully untethered reconfigurable system.

Therefore, we introduce a magneto-thermomechanically triggering actuation to overcome the tethered problem and avoid the need for additional high thermal power and extended time for reprogramming. Our method couples a localized magnetic moment with the prestressing of SMP lattice structures in a fully noncontact triggering manner, eventually providing untethered, reversible, reprogrammable deformations of lattice structures with shape locking. The magnitude and direction of local magnetic torque and thermomechanical force, conceived by interactive bending mechanics and reverse stiffness, enable complex yet versatile transformations.

## Results

### Reversibility with shape locking

Figure 1a illustrates the working principle of reversible transformation of a straight lattice structure to a curved lattice via magnetic and thermal stimuli. Introducing a magnetic force-driven thermomechanical deformation to the conventional training stage of SMP structures provides additional reconfigurability; one can achieve a reversible transformation of SMP structures. We use a Halbach array holding a thermal bath to apply an external thermo-magnetic field, as shown in Figure

1a.

We embed permanent magnets, NdFeB, on the vertices or edges of SMP lattices made of polylactic acid (**PLA**), a typical thermoplastic of fused deposition modeling (**FDM**)-based 3D printing whose glass transition temperature, $T_g$, is ~60°C. We apply a magnetic field **B** on the lattice structures, where the lattices were prestressed without deformation because PLA has a relatively high stiffness at room temperature ($E_L^{PLA} \sim 2.4\ GPa$). However, its stiffness significantly drops after passing $T_g$ ($E_H^{PLA} \sim 3\ MPa$ at ~70°C), as shown in Figure 1b[29,30], and the prestress by magnetic force is released with deformation – Transformation I in Figure 1a. The magnetic torque $\mathbf{M}\ (= \mathbf{m} \times \mathbf{B})$ starts rotating the SMP beams via an interaction of an external magnetic field **B** and magnetization of the NdFeB magnets **m**. Notably, the massive drop in the stiffness of the SMP at $T_H$ enables a fast response[30], overcoming a traditional slow thermomechanical actuation, e.g., the deformation of bi-materials via a thermal expansion mismatch. Cooling down the structure to $T_L$ while maintaining **B** locks the deformed shape with a high stiffness of $\sim 2.4\ GPa$. Notably, the SMP structure can retain the deformed shape even though **B** is removed due to the shape fixity effect of the SMP. Reheating the SMP structure above $T_H$ without the application of **B** recovers the initial configuration of the SMP structure due to the shape memory effect of the SMP, as illustrated in Transformation II in Figure 1a.

Integration of the two transformations — thermomechanical deformation and recovery – yields **a** reversible shape deformation, which has long been an unmet target in the SMP research community. Notably, we utilized the traditional thermomechanical training of the SMP[17,31,32] as an additional shape-changing step via magnetic torque – Transformation I, producing a complete reversible transformation. Note that we achieve the reversible transformation via structural mechanics, not by material synthesis.

We can generate the magneto-thermomechanical deformation of SMP lattices for varying arrangements of magnets on vertices and edges, as shown in Figure 1c – chiral and achiral square lattices. For a lattice's beam member whose length, thickness, and depth are $L$, $w$, and $b$, respectively, we can obtain the beam's pure bending deflection with temperature-dependent stiffness by solving the following equations: (i) the slope of the beam member $\theta = L/2\rho$; (ii) the radius of the curvature $\rho = EI/\tau$, where $E$ is the modulus of the SMP and $I (= bw^3/12)$ is the area moment of inertia of the beam member; (iii) magnetic torques $\boldsymbol{\tau} = \mathbf{m} \times \mathbf{B}$ ($\tau = mB\sin(\pi/2 - \theta)$) on the beam members, where $m$ is the magnitude of the magnetic moment and $B$ is the magnitude of the magnetic field. The deformation of the beam can be quantified with the maximum deflection at the center $y_{max} = \rho(1 - \cos\theta)$ (please find the detail of derivation in Supplementary Note I).

To find the geometric and operational requirements for Transformation I – negligible deformation with prestressing at $T_L$ and noticeable deformation at $T_H$ by the stiffness drop of the SMP and magnetic torque, we can find geometric (e.g., slenderness ratio) and operational (e.g., temperature, magnitude of **B**) conditions. For a beam's length of $50\ mm$, we compare the maximum deflection under magnetic actuation between room temperature ($T_L = 25°C$) and high temperature ($T_H = 80°C$) for the slenderness ratio $w/L = 0.01$–$0.10$ and $|\mathbf{B}| = 20, 50,$ and $80\ mT$. As observed in Figure 1d and Figure S11, a structure with a thick layer ($w/L > 0.02$) exhibits an almost zero deformation ratio $\delta(= y_{max,T_L}/y_{max,T_H})$, indicating a negligible deformation at $T_L$. However, a structure with a thin layer ($w/L < 0.02$) shows an increase in $\delta$ up to 12% under a strong magnetic field ($80\ mT$), implying that a thin structure cannot hold its prestress condition against **B** despite the high modulus of PLA at $T_L$. Therefore, one must carefully design the beam geometry of SMPs for reversible magneto-thermomechanical transformation.

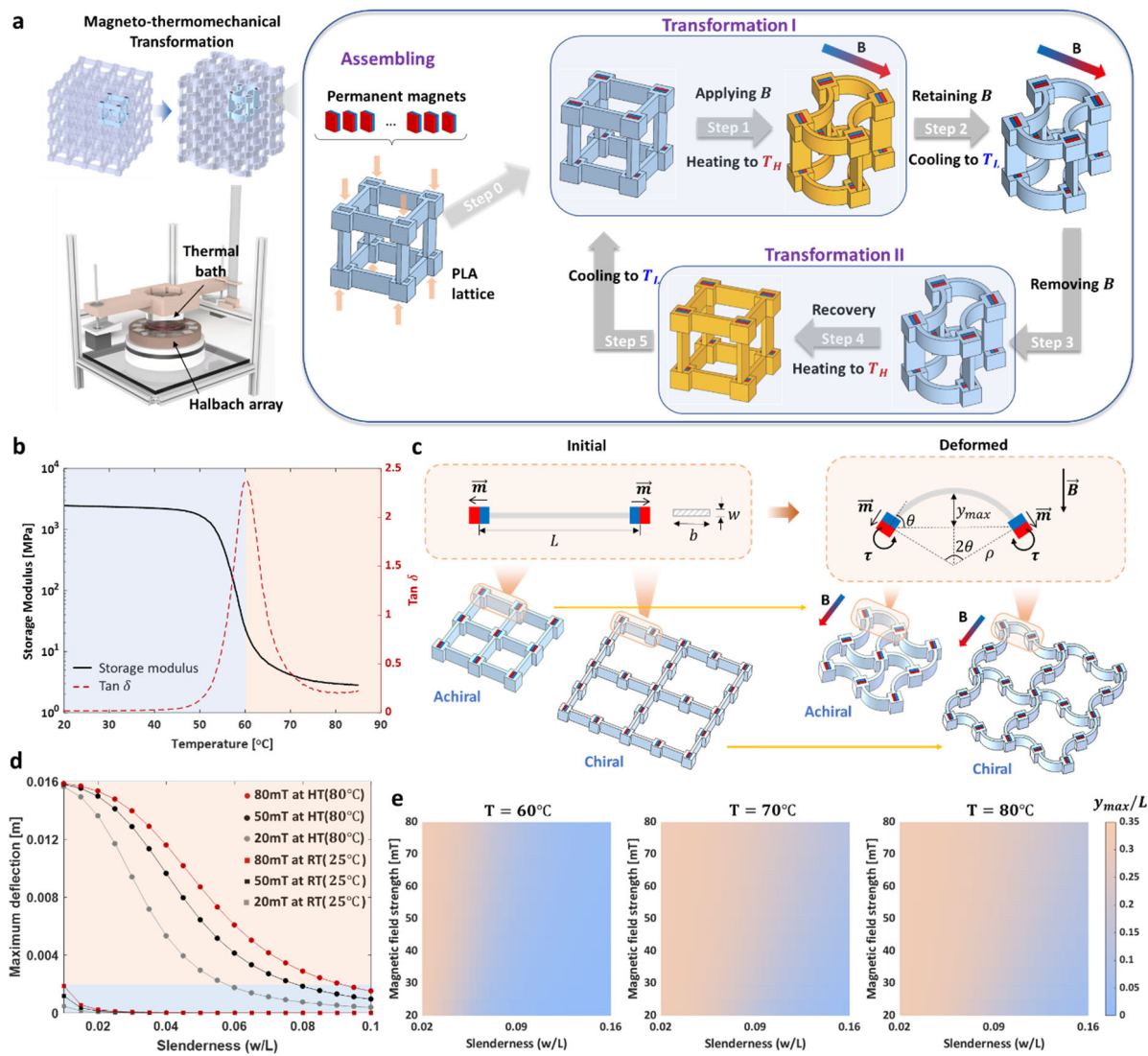

*Figure 1. A magneto-thermomechanically triggered SMP system*: (a) illustration of the working principle. PLA lattices embedding NdFeB permanent magnets transform into predefined shapes and recover to the initial configuration via magnetic field and heat; (b) storage modulus and $T_g$ of PLA obtained by dynamic mechanical analysis (DMA) for varying temperatures; (c) a beam deflection model of lattices by micro-magnetic torque; (d) maximum deflection of a beam for varying $|\mathbf{B}|$ and temperature, indicating the critical slenderness ratio for reversible actuation; (e) normalized maximum deflection for varying slenderness ratios and magnetic field strength for $T = 60℃$, $70℃$, and $80℃$.

We obtain $\mathbf{m}$ of NdFeB from the hysteresis loop measured by a vibrating sample magnetometer in Figure S12. We also measure the magnetic moment versus temperature ($\mathbf{m}$–$T$ curve) to determine the critical working temperature of NdFeB to avoid demagnetization. Figure S13 shows that the magnetization of NdFeB remains at a relatively high value below $60℃$ while slightly losing $\mathbf{m}$ as we increase the temperature. To ensure that the embedded NdFeB can effectively actuate the SMP lattice, we set the acceptable decay ratio of magnetization to be 20%. Therefore, the activation temperature $T_H$ should **not** exceed $80℃$.

A full deformation requires an actuation time of $30$, $10$, and $5\,s$ at $60℃$, $70℃$, and $80℃$, respectively, as demonstrated in Figure S14. We conducted a cyclic magneto-thermomechanical test of an SMP lattice ten times to validate the cyclic repeatability. The lower activation temperature and shorter actuation time are known to provide a better recovery behavior[33]. However, we may select an activation temperature of 70°C rather than 60°C because of the relatively long activation time (30 $s$) at 60°C. The deformed shapes for Transformations I and II at each cycle are presented in Figure S15 and Supplementary Video 1, demonstrating that the reversible transformation is also repeatable.

We investigate the effect of ($w/L$), $T_H$, and $|\mathbf{B}|$ on the deformation. Utilizing the analytical model, we calculate the normalized maximum deflection ($y_{max}/L$) in the ranges $0.02 < w/L < 0.16$ and $20 mT < |\mathbf{B}| < 80 mT$ at three activation temperatures ($T_H = 60°C, 70°C, and\ 80°C$) and construct phase diagrams, as shown in Figure 1e. We also plot the deformed shape of the lattices for varying $|\mathbf{B}|$ in Figure S16, associated with Figure 1e. To verify the phase diagrams in Figure 1e, we conduct an experimental parametric study by printing samples with four slenderness ($= 0.04, 0.06, 0.08, 0.10$) and activating them under $|\mathbf{B}| = 20, 50, and\ 80\ mT\ and\ T_H = 60°C, 70°C, and\ 80°C$). An excellent agreement among the experiment, analytical prediction and finite element simulation (Figures S17) proves the accuracy and reliability of our analytical model for design guidelines of transformation in Figure 1e. The process of finite element simulation is introduced in Supplementary Note II.

**Structural tunability with remote control**

We can design tunable lattices by breaking and relaxing symmetry using the magneto-thermomechanical transformation. Figure 2a displays a set of reconfigurable polygonal lattices: square, triangular, and hexagonal, with various arrangements of hard magnets on the lattices. Under the external magnetic field and thermal stimuli, the micro-torque at the vertices and edges of the straight struts transforms the straight lattices into curved lattices, causing the polygonal lattices to deform into achiral or chiral lattices (see Figure S18 and Supplementary Video 2).

The magneto-thermomechanical transformation can build a functionally graded lattice (**FGL**) for nonhomogeneous design with spatially varied curvatures[34], sequential deformation, and adaptive control of the elastic wave[35]. There are two ways to generate FGL in this work. One is with a nonhomogeneous geometry of lattices applied by a homogeneous local magnetic control. The other is a uniformly distributed lattice geometry applied by a nonhomogeneous local magnetic control. Figure 2b shows an example of FGL constructed by the former method – applying a thickness gradient $\nabla t$ to square lattices with a homogeneous arrangement of hard magnets; we apply $t = 0.5, 0.8,$ and $1.1\ mm$ for thin, medium, and thick regions, respectively, while maintaining a lattice size of $10\ mm$.

We can design an active kirigami structure with shape locking by reducing the beam length to zero and applying a hinge at one corner, as shown in Figure 2c. The fully collapsed configuration can provide extreme stiffness, whereas the fully expanded shape can produce relatively low stiffness. Notably, an extra magnetic force does not need to maintain the extreme configurations due to the shape locking of the SMP, which was not able to be realized with conventional kirigami design with soft materials[36].

The magneto-thermomechanical transformation can vary the magnitude of the curvature, tuning Poisson's ratio and stiffness, as shown in Figures 2d and 2e, respectively. The sudden jump of Poisson's ratio from a positive to a negative value for Lattices B ($\theta = 20°$) and C ($\theta = 30°$) in Figure 2d and the tunable stiffness in Figure 2e remarkably agree with the previous theoretical work[37]. See Figure S2–S3 and Supplementary Note III for further details on the experimental demonstration.

The tunability with the untethered active structures can be used for an elastic wave filtering, as demonstrated in Figure 2f and Figure S19, where the tunable curvature by the magneto-thermomechanical control can generate band gaps for a specific frequency range. Again, the reconfigured structure does not require an extra magnetic force to retain the deformed shape due to the shape locking of the SMPs.

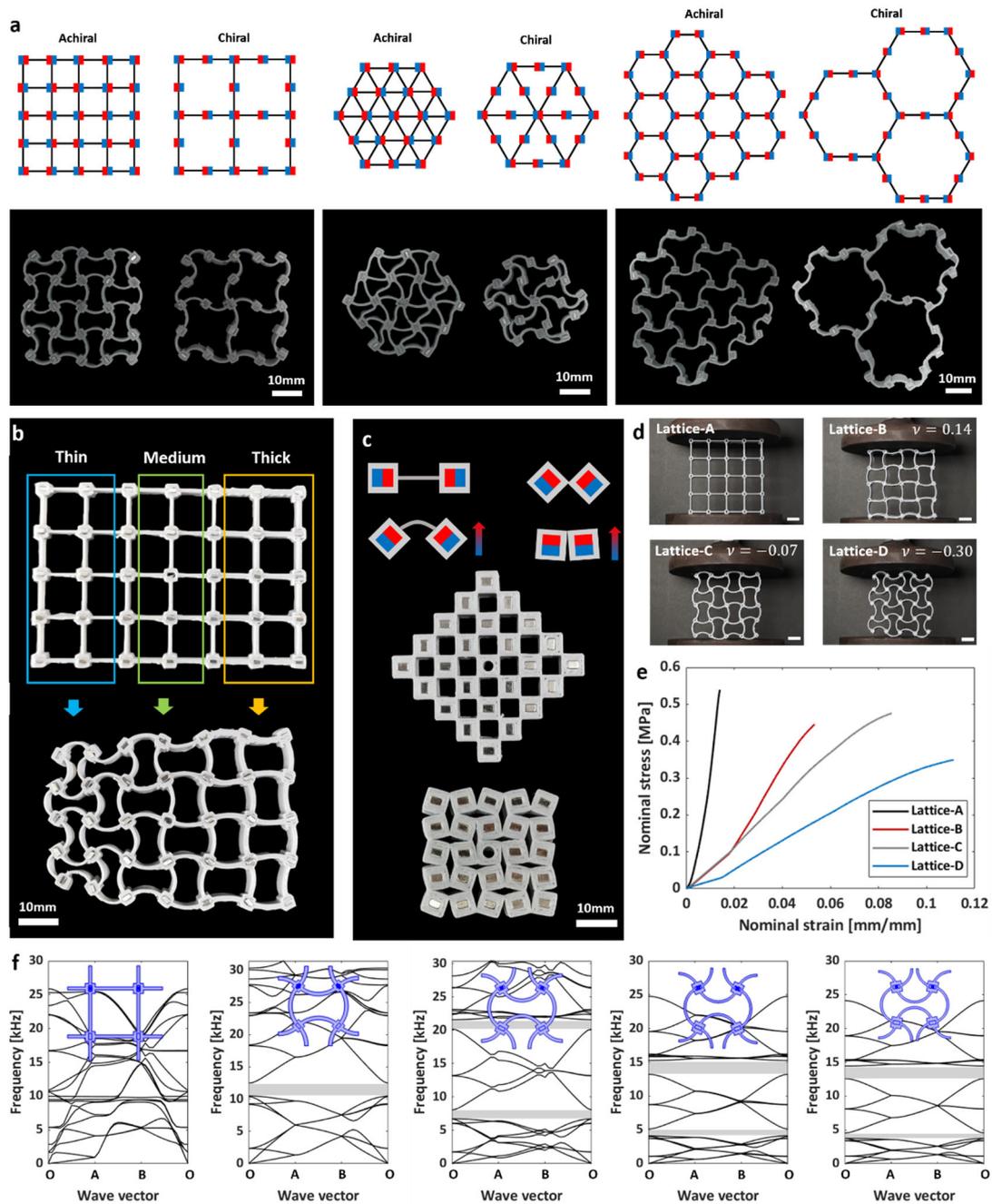

*Figure 2. Design of tunable structures*: (a) square, triangular, and hexagonal lattices with a different arrangement of hard magnets for chiral and achiral deformations; (b) functionally graded square lattice; (c) fully collapsible kirigami structure with shape locking; (d) tunable Poisson's ratio by magneto-thermomechanically deformed curvature; (e) tunable stiffness by magneto-thermomechanically deformed curvature; see Figures S2–S3 and Supplementary Note III for more detailed experimental demonstration; (f) FE simulations of elastic wave filtering with a remotely tunable square lattice.

**Reprogrammable transformations**

Most active metamaterials possess only a single deformation mode after fabrication[6,24,38]. A symmetry-breaking approach generates multimodal motions but still produces limited modes without resetting function[6]. Reprogramming by demagnetizing and reorienting magnetic particles has more potential to increase the transformation modes. Nevertheless, it requires extra energy and extended time for reprogramming, such as laser heating and welding[11,26,27].

This work demonstrates the multimodal and reprogrammable transformation of a single lattice structure by semi-automatic replacement of magnetic arrangement (Supplementary Video 3). As shown in Figure 3a, a square lattice has pockets on the vertices and edges of the beam structures to insert hard magnets selectively. Mounted magnets on the vertices can generate a square achiral pattern (Mode 1) when exposed to a thermo-magnetic field, as shown in Figure 3b. Removing heat to $T_L$ while maintaining the magnetic field, we can make the structure hold the deformed shape. Even upon removing the magnetic field at $T_L$, the system maintains the deformed shape due to the shape locking of the SMP. One can transform the square achiral structure back to the original straight form with an exposure to $T_H$ due to the recovery of the SMP. Similarly, by rearranging hard magnets on the edges (Mode 2) or vertices/edges (Mode 3), we can generate additional reversible transformations – directional bending and square chiral, as shown in Figures 3c and 3d, respectively. Notably, all the transformations can be reset to the straight patterns, reprogrammable to others without additional extreme energy for reprogramming.

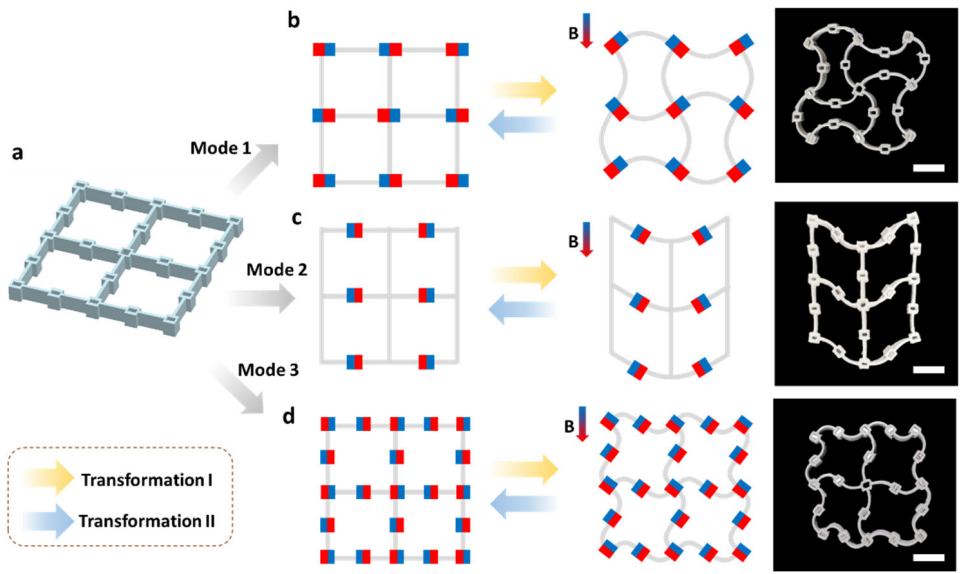

*Figure 3. Multimodal deformation of a square lattice with a different geometric arrangement of hard magnets: (a) initial square lattice with pockets on the vertices and edges; (b) magnets on the vertices generate an achiral pattern – Mode 1; (c) magnets on the edge generate a bending pattern – Mode 2; (d) magnets on the vertices and edges generate a chiral pattern – Mode 3; the scale bar represents 10 mm.*

To realize fully automatic reprogrammable transformations without rearranging magnetic patterns, we use a bistable curved beam[39-41] with an asymmetric magnetic arrangement while changing the direction of a magnetic field **B**, with four selected designs illustrated in Figure 4a.

Design 1 contains a curved beam and two identical magnets ($m_1 = m_2$) at the ends. The magnetic moment is applied along the tangential direction of the beam, pointing toward the center at a downward **B**; both micro-torques on the magnets rotate downwards, making the initial curved shape C1 flip to another stable mode C2, as shown in Figure 4a. One can achieve a reversible transformation with an upward **B**. Theoretically, Design 1 can achieve S-shaped (S1 and S2 in Figure 4a). However, these two shapes are unstable and rugged to accomplish in an experiment.

With symmetry breaking with different magnetization (**m**) of NdFeB, e.g., $m_1 < m_2$ in Design 2 in Figure 4a, we can obtain four deformation modes, including S1 and S2 shapes. Similar to Design 1, the transformation between two C-shapes only requires downward or upward **B**. Rotating **B** by $\theta_s$ can selectively maximize the magnetic torque at one side and minimize it at the other side, providing a double-curved shape (S-shapes). S-shapes (S1 and S2) can transform to C-shapes (C1 and C2) with downward and upward **B**. Notably, we cannot directly transform S1 into S2 and vice versa. Instead, a two-step transformation is required: deformation to a C-shape followed by another S-shape, as the transformation paths are presented in Figure 4a. Adding the third magnet ($m_3$) at the center of the curved beam can result in a direct transformation between the C- and S-shapes, which can be applied in Designs 3 and 4 of Figure 4a.

We can extend the 1D bistable unit's transformation to a 2D unit-cell design. Mapping the beam's deformation by the asymmetric micro-torque for the varying direction of **B** can help us design the transformation of the 2D lattices, e.g., the deformation map for Design 2 in Figure 4b with experimental validation in Figure S20. We can classify the relationship between the net magnetic torque, $\mathbf{m}_{net}(= \mathbf{m}_1 + \mathbf{m}_2)$ and **B** with a symmetry axis connecting the end nodes of a beam in Figure 4b by dividing the magnetic torque into four groups: Q1–Q4 in Figure 4c. Region Q1, where $\mathbf{m}_{net}$ and **B** are in the same quadrant, shows no crucial transformation; the C1 mode of Design 2 remains the same with slight perturbation. The **B** in region Q4 also does not transform the beam's initial shape C1. However, if **B** is oriented in region Q2, the beam's initial shape can partially cross the symmetry axis to build an S-shape or snap through to make a C2 shape. The **B** in region Q3 provides either no transformation or snapping to a C2 mode. One can apply this method for varying polling directions of hard magnets, as shown in Figure 4b. We also developed an analytical model for the bistable curved beam (please see Supplementary Note IV), which indicates the deformed shape of the beam under various directions of magnetic fields. As we can see, the analytical prediction (Figure S4b) shows an excellent match with the proposed deformation map in Figure 4b. This model also helps explain why the symmetry-breaking design can generate S-Shape (Figure S5).

We developed a coding method in grid form to build a lattice with a prescribed transformation. Each grid represents one unit, where the grid color denotes the four groups (Q1–Q4) of **B**. The letter and its color describe the deformation (C and S modes); red, orange, and blue represent deformations entirely, partially, and without crossing the symmetry axis. D design with a combination of 1D bistable units with asymmetric magnetization can multiply the chance of transformation. For example, a four-node circular lattice in Figure 4d can produce six representatives, including chiral, achiral, and mixed chiral–achiral shapes of square lattices for varying directions of **B**. Similarly, we can design a three-node circular lattice in Figure 4e to produce a multimodal transformation.

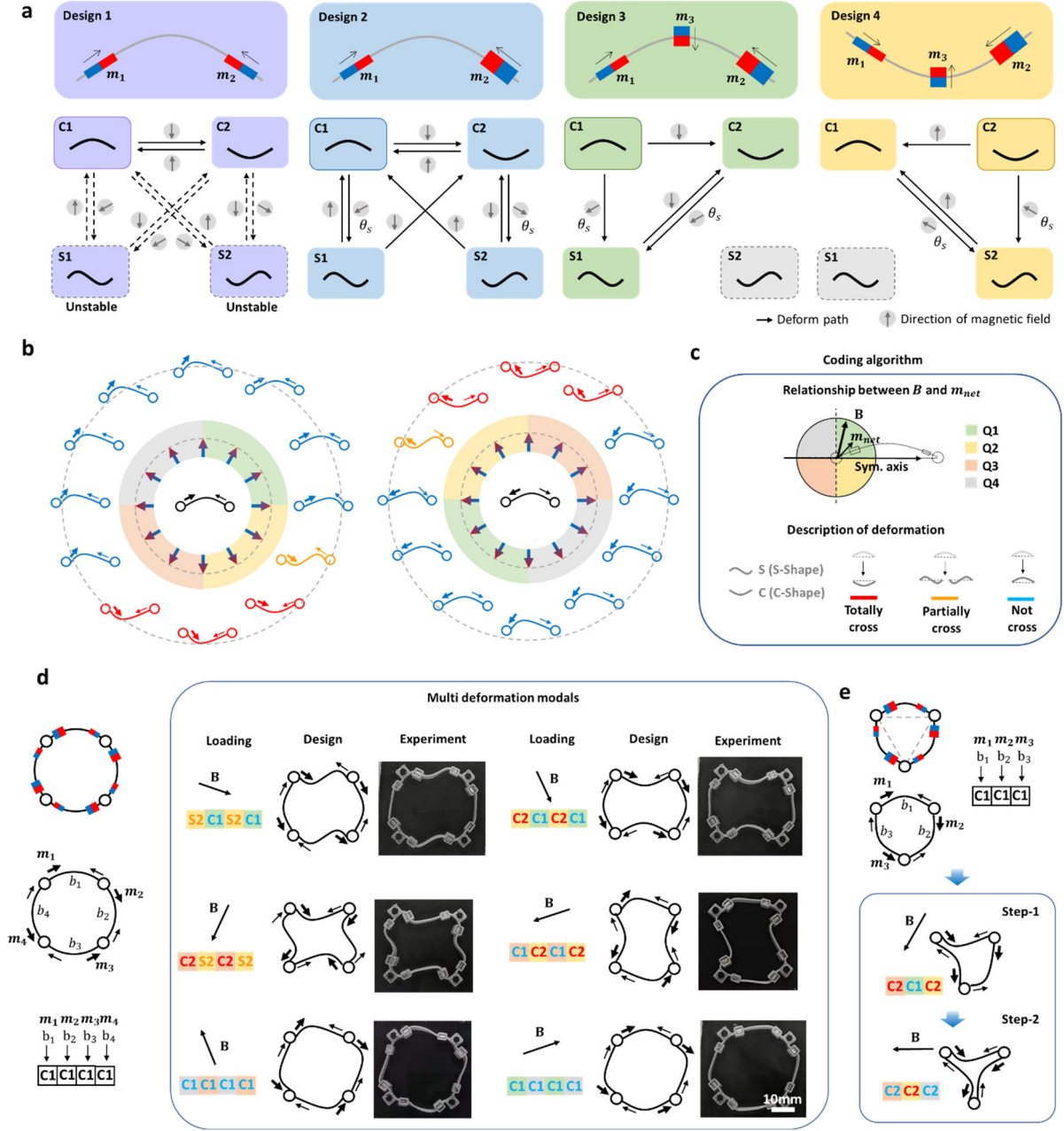

*Figure 4. Multimodal deformation with bistable curved beams and symmetry breaking:* (a) four curved beams with varying magnet arrangements; (b) deformation map of a beam unit in Design-2, indicating its deformed shapes with three modes (blue, red, and yellow) for varying directions of a magnetic field; (c) coding algorithm revealing i) the relationship among the net magnetic torque $m_{net}$, magnetic field $B$, and the symmetry axis; ii) the shapes and positions for deformation modes; (d) a four-node circular lattice with six representative deformation modes for varying directions of $B$; (e) three-node circular lattice with two representative deformation modes.

One can extend the unit cells to $n \times n$ lattices in periodic and non-periodic ways, realizing transformable lattice materials. The coin-like and ginkgo-leaf-like lattices demonstrate a multimodal transformation, as shown in Figures 5a and 5b. The readers can find the magnetic arrangements and all the deformation modes in Figure S21 in the Supplementary Information. Notably, the coding algorithm can help describe the configurations and analyze the tessellation capability for periodic lattice structures.

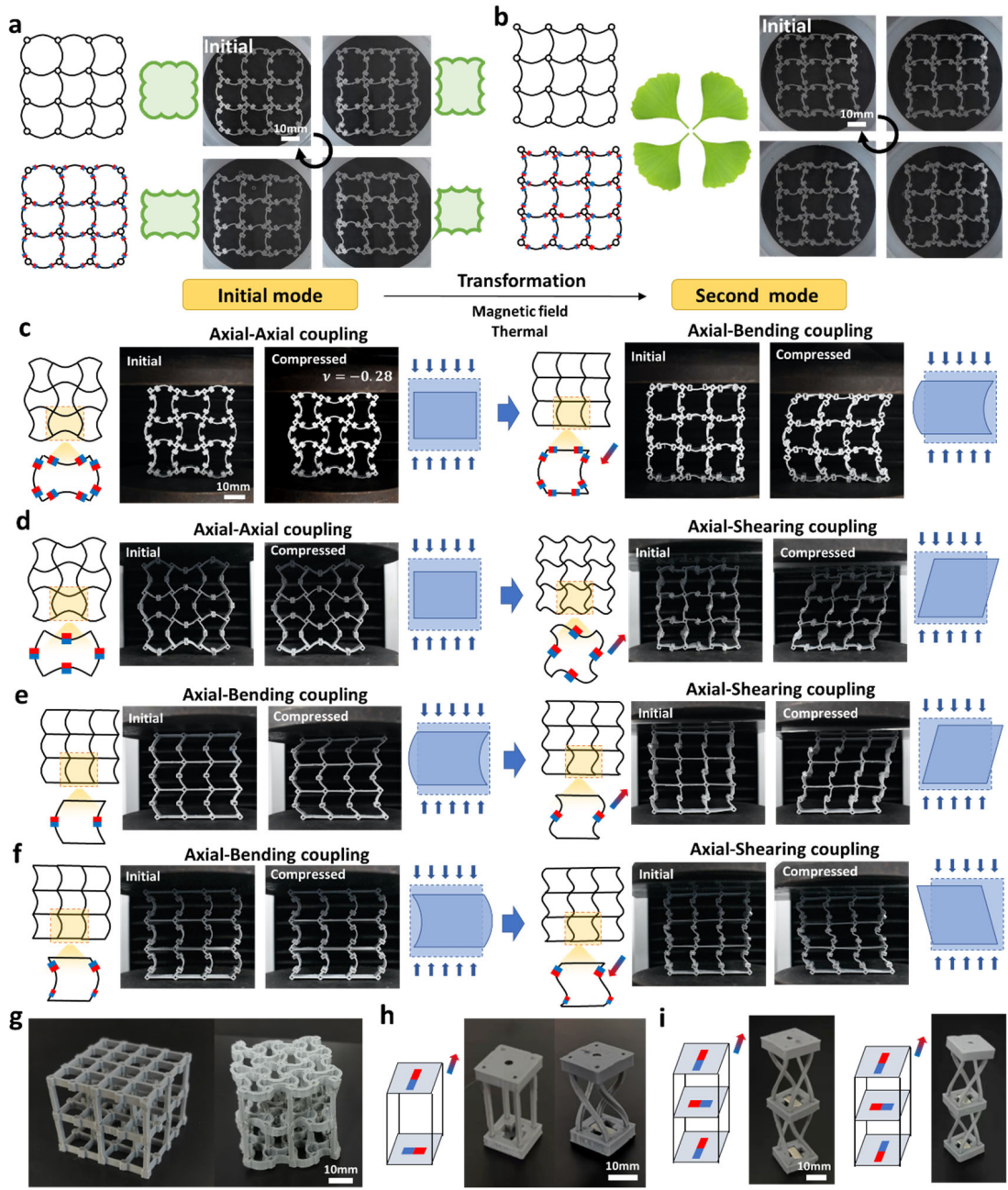

*Figure 5. 2D and 3D lattices with multimodal and multi-coupling effects*: (a) A lattice-like array of coins stacked can experience a loop of deformed sequences. (b) A fish-scaled lattice (point group 1) can behave like ginkgo leaves and serve as the magnetic field indicator. (c–f) Transformations between different coupling effects, including (c) from an axial–axial coupling (also known as negative Poisson's ratio) to axial–bending coupling, (d) from axial–axial coupling to axial–shearing coupling, (e) from axial–bending coupling to axial–shearing coupling, and (f) another design for axial–bending coupling to axial–shearing coupling. (g) 3D lattice with a bending mode transformation, (h) 3D lattice with a twisting mode transformation, and (i) multi-layer design for multi-twisting modes.

Mechanical coupling can provide an interactive design that can communicate with adjacent units[42,43]. Beyond the most well-known mechanical coupling (Poisson's effect), lattice structures can produce anisotropic mechanical couplings such as axial–shear[37], axial–bending[44], and axial–twist couplings[45] by breaking the symmetry of regular lattices. Our magneto-thermomechanical transformation enables a structure to have multiple coupling effects (Supplementary Video 4). A tetra-achiral lattice with a negative Poisson's ratio (axial–axial coupling) can transform into a structure with an axial–bending coupling upon changing the direction of **B**, as shown in Figure 5c. Similarly, the tetra-achiral lattice can also change its shape for an axial–shear coupling, as shown in Figure 5d. The transformation between structures with axial–bending and axial–shear couplings is possible, as we demonstrate multi-coupling effects in Figures 5e and 5f.

We can extend our magneto-thermomechanical transformation to 3D lattices potentially applicable in tuning static and dynamic physical properties. Figure 5g shows a 3D structure conceived by vertical stacking of the in-plane chiral transformation. Notably, the rotation of hard magnets at nodes leads to the bending of straight beams. The transformed structure has an axial–shear coupling in the in-plane direction[37]. Figure 5h shows a structure where hard magnets are placed on the top and bottom plates with different polling directions, producing twisting of vertical columns after the magneto-thermomechanical transformation. The transformed structure has an axial–twisting coupling effect[45]. Stacking additional units in the vertical direction can produce a multiple twisting transformation mode, as shown in Figure 5i, that can also be used for tuning the magnetite of the axial–twisting effect; the same or opposite directional polling of the hard magnets on the top and bottom plates double or cancel the axial–twisting effects, respectively. Notably, it is challenging to directly print the twisted structure without supporting materials, especially in extrusion-based 3D printing such as fused deposition modeling (FDM) and direct ink writing (DIW). However, our transformation method can provide a solution.

## Discussion

Despite its shape-shifting with large deformation, the irreversibility of SMPs has long been an intrinsic obstacle to realizing reversible shape morphing structures[20]. On the other hand, despite fast and remote actuation capability, magnetic actuation's inherent disadvantage – the need for a magnetic force to hold a transformed shape has been a critical bottleneck for moving forward with future reconfigurable structures[1,6].

By combining two physics concepts — prestressing by magnetic control and thermomechanical training/recovery of the SMP — we can provide mutual assistance for advanced active metamaterials without requiring new materials synthesis; this work can break through the bottlenecks of SMPs and magnetic control by allowing reversible control of SMPs without the requirement of magnetic force for shape-holding after deployment. In addition, another structural mechanics-inspired design approach using instability and recovery of the SMP can enable a reprogrammable transformation without extra energy for reprogramming. A comparison of property and performance between our design and other common smart materials is listed in Supplementary Table 1.

Previously, most SMP structures were trained by global pre-strains from mold and external mechanical loads[46,47]. Our deformation capability with local magnetic torque can reconfigure with complex geometry such as functionally graded structures in Figure 2b, which could not build with the traditional training method of SMPs[17,31,32]. It is also highly challenging to hold a buckled shape of soft structures without releasing external loadings[46,47]; however, the shape fixity of SMPs can resolve this issue in this work. Moreover, it is challenging to generate a higher mode (a low-wavelength mode) deformation of unstable structures[46,47]. Our asymmetric magnetization arrangement and instability design can generate a higher mode deformation, as demonstrated in Figure 4a.

Even though we demonstrate the programmability and reprogrammability of our structures by embedding permanent magnets on the SMP lattices printed by FDM, the principle in this work can be applied to direct ink writing (DIW) while the magnetization and fabrication process can be coupled[1] for a selected SMP matrix, e.g., thermoset polymers[14] or solvent-cast thermoplastic polymers[48]. A high-resolution printing without the assembling process enables the scaled-down of this triggering mechanism, which can greatly expand the application scenarios, e.g., in biomedicine devices or micro-robots. The fabrication methods and corresponding applications of the proposed active materials on various scales are shown in Figure S22.

Unlike previous reprogramming methods, which require high-power heating energy, e.g., above the Curie temperature ($118°C$ for chromium dioxide $CrO_2$)[11] or laser power (0.06 W)[26], our reprogramming method for multimodal transformation does not require high energy but only heating to $\sim 60°C$ to reset with a stress-free condition.

Our multi-physics-based approach represents a paradigm shift in the design of reconfigurable structures by primarily using structural mechanics without the need for new materials synthesis or extra energy for reprogramming. This structural design opens a new branch to construct future robotic matter with reversible, shape-holding, untethered, and energy-efficient reprogrammable functionality.

Our magneto-thermomechanics-inspired design can vastly advance the development of future active metamaterials for reconfigurable structural applications requiring fast, untethered, reversible, and reprogrammable (multimodal) transformability with shape locking. The multi-physics approach in this work to design reconfigurable structures can be used for active metamaterials with tunable properties in static and dynamic manners in actuators, biomedical and wearable devices, flexible electronics, and aerospace and soft robotic applications (see Supplementary Table 2).

## Materials and Methods

### Fabrication of the structure

All the SMP structures were fabricated using a fused deposition modeling (FDM) 3D printer (Ultimaker 2+, Ultimaker, The Netherlands) using polylactic acid (PLA) filaments (PolyLite PLA, Polymaker). Neodymium–iron–boron (NdFeB) permanent magnets (N35, Shenzhen Lala Magnet Co. Ltd., China) were inserted into the PLA lattice to serve as the actuator component in the SMP–Magnet system.

### Dynamic mechanical analysis (DMA) test of PLA

A DMA test was conducted on 3D printed PLA strips ($30 \times 4 \times 1 mm^3$) to measure its viscoelastic properties such as the glass-transition temperature and storage modulus. A DMA machine (DMA Q850, TA Instruments, USA) with a tension clamp was used to perform the test. The samples were tested in a temperature ramp mode over a temperature range of $20°C$–$85°C$ at a heating rate of $3°C/min$ and a frequency of $1\ Hz$.

### Magnetic characterization of permanent magnets

The magnetic moment density (magnetization) versus magnetic field strength ($M$–$H$ curve) or temperature ($M$–$T$ curve) of the NdFeB permanent magnet was measured with a vibrating sample magnetometer (VSM; MPMS3, Quantum Design, USA). For the M–$H$ curve, a permanent magnet with the size $3 \times 2 \times 1 mm^3$ was tested under a sweep of external magnetic fields ranging from 3T to -3T. The hysteresis loop was obtained, from which the magnetic moment density of NdFeB was calculated by dividing the magnetic moment when the external magnetic field was zero to the sample volume. The experiment for the M–$T$ curve was conducted with a permanent-magnet sample with the same size in the temperature range from 25℃ to 120℃ under an external magnetic field of 100 $mT$. The decay ratio of magnetization can be calculated using $\eta = |M - M_{25℃}|/M_{25℃} \times 100\%$.

**External actuation**

Samples were thermally triggered in a glass of hot water at the desired temperature. The hot water for deformation was acquired from a tank of hot water source whose temperature was controlled by an immersion heater (Anova, USA). The external magnetic field was generated by a horizontally placed Halbach array consisting of twelve cubic NdFeB permanent magnets (N35, $25 \times 25 \times 25 mm^3$) as discuss in Supplementary Note V.

The direction of the magnetic field was adjusted by rotational motion of a Halbach array (Supplementary Note VI), while the magnitude of the magnetic field intensity was controlled by altering the vertical distance of specimens from the Halbach array. Both the rotational motion of the Halbach array and the vertical motion of the specimen could be programmed before the test and instantly adjusted during the test from a connected computer. These methods resulted in a tunable magnetic field intensity within a range from 20 to 80 mT, with magnetic field availability in all the in-plane directions of the horizontal planes at different heights. The relationship between the specimen height and magnetic field intensity and the magnetic-field-intensity distribution at horizontal planes of different heights are further discussed in the supplementary information.

## Acknowledgments

This research is supported by the Shanghai NSF (Award # 17ZR1414700) and the Research Incentive Program of Recruited Non-Chinese Foreign Faculty by Shanghai Jiao Tong University. We thank all members of S-Lab at UM-SJTU Joint Institute for their comments.

## Author contribution

J.J. designed and supervised the research; B.Z. designed the structures, built the analytical model, performed the experiments, and analyzed the data. Z.L. constructed the experimental setup, performed the mechanical testing, and analyzed the experimental data. Z.C. designed the mechanical coupling structures. K.X. performed the acoustic analysis of the lattices. S.S. helped construct the magnetic actuation setup. B.Z., Z.L., Z.C., S.S. and J.J. wrote the manuscript.

## Competing interests

The authors declare that they have no competing interests.

## Data availability

All data needed to evaluate the conclusions in this work are present in the paper and the Supplementary Information. Additional data related to this paper may be requested from the authors.